\documentclass[manuscript,table,authorversion,nonacm]{acmart}

\usepackage{natbib}
\usepackage{multirow}
\usepackage{appendix}

\newcommand{\janwillem}{L1{}}
\newcommand{\michel}{L2{}}
\newcommand{\mirjam}{L3{}}
\newcommand{\lennie}{L4{}}
\newcommand{\susanne}{B1{}}
\newcommand{\mezen}{B2{}}
\newcommand{\jos}{B3{}}
\newcommand{\bob}{B4{}}
\newcommand{\tycho}{B5{}}
\newcommand{\arno}{B6{}}
\newcommand{\tom}{N1{}}
\newcommand{\gerard}{N2{}}

\copyrightyear{2024} 
\acmYear{2024} 
\setcopyright{rightsretained} 
\acmConference[FAccT '24]{The 2024 ACM Conference on Fairness, Accountability, and Transparency}{June 3--6, 2024}{Rio de Janeiro, Brazil}
\acmBooktitle{The 2024 ACM Conference on Fairness, Accountability, and Transparency (FAccT '24), June 3--6, 2024, Rio de Janeiro, Brazil}\acmDOI{10.1145/3630106.3658926}
\acmISBN{979-8-4007-0450-5/24/06}

\begin{document}


\title[Diversity of What?]{Diversity of What? On the Different Conceptualizations of Diversity in Recommender Systems}

\author{Sanne Vrijenhoek}
\email{s.vrijenhoek@uva.nl}
\orcid{0000-0002-1031-4746}
\affiliation{
    \institution{AI, Media \& Democracy Lab and Institute for Information Law, University of Amsterdam}
    \city{Amsterdam}
    \country{The Netherlands}
}

\author{Savvina Daniil}
\email{s.daniil@cwi.nl}
\orcid{0000-0001-8888-2869}
\affiliation{
    \institution{Centrum Wiskunde \& Informatica}
    \city{Amsterdam}
    \country{The Netherlands}
}

\author{Jorden Sandel}
\email{j.sandel@uva.nl}
\orcid{0009-0008-3571-5125}
\affiliation{
    \institution{University of Amsterdam}
    \city{Amsterdam}
    \country{The Netherlands}
}

\author{Laura Hollink}
\email{l.hollink@cwi.nl}
\orcid{0000-0002-6865-0021}
\affiliation{
    \institution{Centrum Wiskunde \& Informatica}
    \city{Amsterdam}
    \country{The Netherlands}
}

\renewcommand{\shortauthors}{Vrijenhoek, Daniil et al.}

\begin{abstract}
Diversity is a 
commonly known principle in the design of recommender systems, but also 
ambiguous in its conceptualization. Through semi-structured interviews we explore how practitioners at three different public service media organizations in the Netherlands conceptualize diversity within the scope of their recommender systems. We provide an overview of the goals that they have with diversity in their systems, which aspects are relevant, and how recommendations should be diversified. 
We show that even within this limited domain, conceptualization of diversity greatly varies, and argue that it is unlikely that a standardized conceptualization will be achieved. 
Instead, we should focus on effective communication of what diversity in this particular system means, thus allowing for operationalizations of diversity that are capable of expressing the nuances and requirements of that particular domain.     
\end{abstract}



\maketitle

\section{Introduction}
\label{sec:introduction}
The concept of diversity is at the same time omnipresent and very ambiguous~\cite{fazelpour2022diversity}. 
In popular discourse, diversity usually refers to the variation of human characteristics, often related to a notion of identity politics~\cite{bernstein2005identity}; in biological research, as a qualifier to the health of an ecosystem~\cite{van2019diversity}; and in media studies, as a 
concept adjacent to pluralism, expressing whether a news selection contains a plurality of sources, voices and opinions~\cite{karppinen2013rethinking}. While all interpretations are valid and intuitively seem to reflect similar concepts, they differ in their operationalization in a way that is unique to their domain. At the same time, diversity is consistently noted as a social value that is beneficial to pursue, though in different capacities. 
\\
\\
In the recommender systems domain accounting for the diversity of a recommendation can help avoid monotony \cite{zhangavoiding2008,castells2021novelty}. This follows from the assumption that a recommender system that is purely optimized on predicted relevance will result in a feedback loop and thus prioritize similar content, leading to `more of the same'. There is as such also a business case to be made for diversity. While it has been challenging to show empirically, diversity may lead to higher user satisfaction and retention, increasing revenue in the process~\cite{ekstrand2014user,jannach2019measuring}. This argument can be extended for the news domain, where worries over filter bubbles that reinforce existing beliefs by only showing content that align with a user's preferences are especially prevalent~\cite{zuiderveen2016should,michiels2023should}. Having a news recommender system that  pursues diversity could help expose users to things different than they are used to or expect seeing, and tailor to their specific information needs \cite{helberger2018exposure,helberger2019on}. For machine learning in general, it is also important to account for diversity in a social context. Since many algorithms are trained on datasets that are not representative of all groups in society, not accounting for diversity may further ``amplify stereotypes, alienate users, and further entrench rigid social expectations" \cite{mitchell2020diversity}. 
\\
\\
The many different interpretations of diversity are a fundamental challenge to the practical development of recommender systems. Evaluating the performance of a recommender system requires objectively measurable properties, and the field strives to do so in a standardized manner~\cite{zhu2022bars}. Standardization allows for comparability and reproducibility of results and algorithms\footnote{Recent strands of research have noted that this perceived notion of standardization is often false. Even with commonly accepted principles there are often significant differences in implementation, and thus output~\cite{tamm2021quality,shehzad2023everyone}}, and is achieved through, among others, the use of benchmark datasets, sampling techniques and evaluation metrics such as MRR or NDCG~\cite{zangerle2022evaluating}, but has not yet been achieved for diversity.
\citet{loecherbach2020unified}, who did a literature review of existing measures of diversity in media studies, note that ``there is little to no overlap between concepts and operationalizations used in the different fields interested in media diversity''. \citet{lawrence2020trouble}, executing a conceptual analysis of the term `diverse books', notes that the topic is inherently political, and that ``we have yet to arrive at a clear consensus on what the modifier `diverse' means in this and other instances". 

The conceptual unclarity around diversity makes that not only may we implement it differently, we may also \emph{mean} something completely different when we use the term. One could argue that diversity is, in fact, an essentially contested concept~\cite{gallie1955essentially,collier2006essentially}, meaning that it is open for discussion and debate and that we are unlikely to reach consensus on its meaning. As such, striving for agreement or a clear definition may lead to a standstill and hinder progress, as it is unclear what a good operationalization or implementation would look like. It may cause the conceptualization of diversity to be driven by the information that is available or more easily suitable for quantification, rather than what is desirable; furthermore, blindly trusting `objective' measures ``may obscure the fact that the conceptualization of diversity is, eventually, a normative choice" \cite{karppinen2015limits}. Instead, what we need may be a more flexible operationalization that is capable of reflecting the nuances and requirements of the domain it is deployed in.
\\
\\
The aim of this paper is to explore the dimensions in which industry practitioners within a limited domain conceptualize diversity. To this end we conduct a series of semi-structured interviews with practitioners from three different public service media organizations in the Netherlands: a broadcaster, a news organization, and a library. The choice of these organizations is deliberate; though the medium through which they do it differs, they all play an active and important role in the dissemination and curation of information and ideas. As such, we expect there to be a fairly established conceptualization of diversity within the organization, and comparatively more overlap between them than with, for example, a commercial music recommendation service. 

Through analysis of the interviews, and guided by the `components' of diversity formalization defined in \citet{vrijenhoek2022RADio}, we aim to answer the following questions: what \emph{goals} do the organizations aim to achieve with the recommendations, which \emph{aspects} do they consider relevant for diversity, and through which \emph{tactics} are the recommendations diversified.
We find that even within the limited domain of public service media recommendation, 
there is a wide variety of conceptualizations of diversity. 
With this, we underline that a standardized definition of diversity in recommender systems is likely not achievable. 
However, this empirical categorisation, albeit non-exhaustive, can assist in the process of conceptualizing and implementing diversity in a particular concrete context.

\section{Related work}


Recommender systems have long moved from evaluating recommendations solely based on accuracy-related metrics. 
Other metrics such as novelty, serendipity and diversity have become a common 
part of evaluation practices \cite{castells2021novelty}. 
Diversity in recommendation is a current topic, and multiple diversification methods have been proposed in recent years \cite{kunaver2017diversity,friedman2023vendi}.
In recommender systems research, diversity is often viewed as the opposite of similarity; a list of recommended items is diverse if the items are sufficiently different between them along a set of axes \cite{10.1145/1060745.1060754}.
However, \citet{jesse2023intra} point to a gap between human perceptions on diversity and intra-list similarity (ILS) commonly used to assess diversity in offline recommender systems experiments.
Through a user study, they find that while ILS can be a good proxy, the details of the implementation matter and require validation in a given domain and application.

When diversity is seen as a social value that needs to be incorporated in a system, standardizing its definition and operationalization becomes even more challenging.
\citet{mitchell2020diversity} define diversity in a subset selection task (e.g., the construction of a list of items to recommend) as ``variety in the representation of individuals in an instance or set of instances, with respect to sociopolitical power differentials (gender, race, etc.)".
As such, they view diversity as a concept with inherently sociopolitical connotations in contrast to the more general term `heterogeneity'.

In the context of media recommendations, diversity is often closely associated with pluralism.
According to \citet{karppinen2015limits}, the interpretation of pluralism and diversity in media is dependent on the political and normative understanding of the role of media in society.
Additionally, defining media diversity is further complicated by its often contradictory aspects and interpretations.
\citet{van1999competition} point out an antithesis between the normative media diversity frameworks of \textit{reflection} and \textit{openness}; one requires content distribution to proportionally reflect societal distributions of relevance, while the other corresponds to perfectly equal representation and attention to all people and ideas in society.
Similar contradictions have been laid out in the domain of library and information studies \cite{lawrence2020trouble}. 
Based on a systematic literature review on media diversity, \citet{loecherbach2020unified}, also referenced in the introduction on the little overlap between different fields' conceptualization of diversity, conclude that relevant research should be guided by an interdisciplinary effort to define and benchmark the concept of media diversity, along with its different sub-dimensions. 
Other work studying diversity in recommender systems would typically acknowledge the complexity of the concept, yet model it following existing technical standards; either as a distance to a user's reading history \cite{harambam2019designing}, political leaning \cite{heitz2022benefits} or as pair-wise distance between the items in a recommendation, for example in topics, categories and/or tone \cite{moller2020not}.
In this work, we aim to contribute to this effort by attempting to map the dimensions of diversity in media recommender systems.

\subsection*{Diversity in Public Service Media}
Diversity in recommendation is especially relevant for public service media (PSM), whose role and societal responsibilities call for a careful consideration of the content they produce and give exposure to.
Many media organizations underline the potential of recommender systems and the importance of reflecting editorial values such as diversity therein~\cite{kruse2023creating,grun2021challenges,boididou2021building,moller2022recommended}.
PSM are often required to offer diverse content, which 
might be at odds with 
the primarily commercial use of recommender systems that mainly intend to increase media consumption, and therefore, profit \cite{hilden2022public}.
Even if we assume societal agreement on the importance of providing diverse content for PSM, the practical implications are harder to lay out. 
\citet{helberger2019on} outlines four models for the normative conceptualization of diversity in recommender systems that serve different purposes: the liberal, the deliberative, the participatory, and the critical.
While each of the models is in its own way relevant for PSM, the work suggests that the critical  perspective, which focuses on the visibility of minority voices that are often disadvantaged in public platforms, is less likely to be encountered in commercial applications, and therefore could be partly served by PSM. 
Regardless, few have succeeded in concrete implementations that reflect diversity as a normative value, in part due to the gap between journalistic values and recommender system evaluation metrics \cite{vrijenhoek2021recommenders}. \citet{moller2023designing} suggests that ``the abstract nature of journalistic values make them hard to account for computationally", and that ``[h]uman journalists have an important role to play in these processes not only to help conceptualise the values themselves but also as part of new algorithmic news practices."
Translating values into concrete algorithmic practices is thus as challenging as it is necessary.

To bridge the gap between theory and practice, the perspectives of practitioners in a given domain can assist, orient, and ground research.
On that note, researchers have conducted interviews with practitioners on the interaction between emerging algorithmic systems and norms and values.
\citet{sorensen2019public} interviewed developers, data scientists, and project leaders from nine European PSM organizations on the topic of implementing recommender systems.
They report that, while interviewees believe diversity to be an essential aspect of their catalogue, they are reluctant to depend on an algorithmic implementation instead of the traditional editorial control.
Additionally, they attribute the general lack of formal definition of diversity to the different understanding between politicians, practitioners, and users.
\citet{bastian2021safeguarding} interviewed practitioners from different departments of two newspapers regarding the impact of algorithmic news recommenders on their organization's norms and mission, as well as how to integrate them in the design of news recommenders. 
They found that, while the interviewees attach varying degrees of importance to different values, diversity is perceived as one of the core values for news recommender design and implementation.
During the interviews we conducted, we specifically focused on the conception and implementation of diversity, which allowed for an in-depth outline of the perspectives and practices of the interviewed professionals.

\section{Method}

For this study we conducted a series of semi-structured interviews with three public service media organizations in the Netherlands: a broadcaster, a news organization, and a library. 
Interviews were carried out in-person and on site in the offices of the interviewees, and took place over a span of four months, between December 2022 and March 2023. At this time each of these organizations were, at different stages of completion, (exploring the possibility of) developing a recommender system to effectively serve content to their users, which also means that decisions about incorporating diversity in the recommendations were actively being made. 

\subsection{Candidate Selection}
Potential candidates were approached through a snowball sampling technique: after initial contacts with the organization were established, interviewees were asked for recommendations of colleagues to interview in the next round. We actively tried to find a set of participants that reflected the composition of the organisations and the relevant figures in the design, deployment and use of the recommender system. 
This process yielded twelve participants in total: six at the broadcaster, four at the library, and two at the news organization. Following \citet{smets2022we}, our goal was to have a good spread of different types of stakeholders, and sought participants with roles related to Business (four participants), Curation (two participants), Product owner (three participants) and Technology developer (three participants). During the snowball sampling we would explicitly ask participants whether they knew of potential interviewees with a role we had not covered yet. Finding all different roles did not always succeed, and we acknowledge this as a limitation of our work. This is also why we are adamant about not making normative claims about the definition of diversity per domain, but rather present it as an exploratory study. Participants were predominantly male (9 out of 12) and white (11 out of 12). This is reflective of the workforce but a caveat for generalization, further outlined in Section \ref{sec:limitations}. Disclosing too many details about individual participants and their roles would undermine the promised anonymity, and thus each participant is assigned a code reflecting only their organization.


\subsection{Interview structure}
The interviews consisted of four parts. In the first part, interviewees introduced themselves and their role within the organization. In the second part interviewees were asked about their general conceptions of diversity, and how these conceptions related to their organization. Here, the goal was to understand the mission of the organization in question, and the role diversity plays in that mission. In the third part interviewees described their knowledge of the recommender systems that were in use by their organization, either in planning or in production. This served as a check that participants were sufficiently informed to be included in the analysis, and to prime the interviewees for the fourth part, in which they were specifically asked about the role of diversity in the recommender systems of their respective organization. The aim here was to go beyond what currently exists, and instead focus on what the recommender system should look like in an ideal situation. 
This part of the interview also contained a small experiment, in which the interviewees were asked to take on the role of a recommender system and rank a set of items while keeping diversity in mind. During the experiment participants were asked to voice their thought process by thinking out loud \cite{charters2003use}. Our primary interest was not the final recommendation generated, but which aspects of the items participants considered before (not) including an item. For each organization we prepared a set of 15 candidate items based on the organization's (potential) catalogue, and included the metadata a user would see when interacting with the system. We attempted to ensure (to the best of our abilities) that there would be enough diversity in this candidate list present. To cover our own blind spots we would ask participants at the end of the experiment whether there was a type of content that they were missing. We acknowledge that our own conception here steers the type of diversity that could potentially be found (see also Section \ref{sec:limitations}). 

For each of the organizations, the metadata always included the items' title, summary and a cover photo. For the library, we also included the authors' name and a set of keywords about the book; for the broadcaster, the title and description of the series the item was part of (if applicable); and for the news organization, the time of publication and the first few paragraphs of the news item. Based on this candidate set, we asked participants to make a diverse selection of 5 items. In order to not influence people's thought processes in what could potentially be relevant aspects of diversity, we deliberately left instructions vague, and did not include a profile of a user to create a recommendation for in the instructions. Instead, we considered the participants' questions for clarification as part of what they deemed relevant for diverse recommendation. 

\subsection{Coding and analysis}
We executed an inductive thematic analysis on each of the interviews, guided by the research questions posed in Section \ref{sec:introduction}: what the organizations aimed to achieve with a (diverse) recommendation, which aspects of the items they would consider during recommendation, and 
which tactics they would employ to diversify the recommendation. The first two authors created a coding scheme based on half of the interviews, which after completion were discussed and merged. With the resulting coding scheme, the authors annotated the half of the interviews they had not previously seen, and extended the coding scheme when gaps were identified. 
Additionally, after the respective coding schemes were merged, the two authors independently coded the same interview and proceeded to compare their outputs. This step helped ensure practical consensus, as the authors were able to align on the specific nuanced interpretation of each of the codes, as well as how it applies in the context of the interviews.
Based on the resulting coding scheme, we created a diversity `map' of relevant aspects and how they relate to each other, which will be discussed in the next section.
The coding scheme is included in Appendix \ref{img:mentions}. After the first draft of the paper was finished, we reached out to all participants. We shared with them which quotes had been attributed to them, and asked them to reach out in case of misunderstandings.

\subsection{Limitations}
\label{sec:limitations}
There are a number of important caveats to consider in light of this method and experiment. By presenting the participants with a list of items to recommend, we inadvertently steer the type of diversity they are likely to mention. For example, if our sample did not contain any mention of politics, the participants might also be less likely to mention this as a relevant aspect. 
Simultaneously, participants may be influenced by what is commonly considered a relevant aspect of diversity, or what interpretations are currently feasible rather than ideal. As such, we cannot draw conclusions on the importance of a certain aspect. 

Another difficulty when running this experiment was accounting for the recency of items, which is extremely relevant for the news organization and the broadcaster, where the first will never want to recommend old news, but the latter may sometimes want to include older content. This was especially an issue given that interviews took place on different days, sometimes weeks apart. For the broadcaster we mixed a stable set of older content with content that was at the day of the interview popular in the system. This has the important caveat that the results between broadcaster interviewees are not fully comparable. For the news organization, we opted for a fixed date a few months in the past. While the interviews are comparable in this case, there is a risk that important contexts are forgotten or mixed up with current events. 

Lastly, while diversity in and of itself is already a complicated and multi-faceted concept, the concepts related to it are too. People may talk about `different ethnicities', and can mean, among others, different skin colors, nationalities or cultures. 
By extension, our findings are largely influenced by the people that participated in the interviews. Organizations are not a monolith, and many of the interviewees were very explicit about not representing the opinion of every member of the organization they belong to. Furthermore, many of the responses are influenced by the background of the participants themselves, and the fact that this research was conducted in the Netherlands. The Netherlands has a strong public service media system with a focus on representing different groups in society~\cite{daalmeijer2004public}. That being said, participants were overwhelmingly white (11 out of 12) and male (9 out of 12). While this is representative of the general demographic working on recommender systems in the Netherlands, people are not as acutely aware of the needs of groups they are not, themselves, a part of~\cite{mcdonald2020intersectional,birhane2022forgotten}. As such, this is a suitable group for a descriptive study (`what is') into the conceptualizations of diverse recommendations in public service media organisations; however, a complete normative formalization of diversity (`what should be') would require the participation of people from different backgrounds and perspectives in order to provide a complete overview.
All limitations considered, the results of this study cannot be used as a definition of diversity. Rather, we aim to show that, even within this limited domain, `diversity' indeed may refer to a wide variety of things. 

\section{Results}
\label{sec:results}
We expect that how organizations operationalize diversity is dependent on what they aim to achieve with their recommendation. We first identify this goal in the following section (Section \ref{sec:goal}). Then, we analyze both the different aspects (Section \ref{sec:aspects}) and tactics (Section \ref{sec:tactics}) mentioned by the participants. 

\subsection{Goal of recommendation}
\label{sec:goal}

Being a public organization means that the primary 
goal of the organization is to provide services that benefit society~\cite{nissen2006public}. As such, each organization's goals with building a recommender system extend beyond selling ads and optimizing for clicks, as is the norm for most commercial organizations. This difference between being a public or a commercial organization is explicitly mentioned by most interviewees (\michel,\mirjam,\lennie,\mezen,\jos,\bob,\tycho,\tom,\gerard), and awareness of this distinction can be considered  central in day-to-day operations. In the absence of optimizing for monetary gains, the goal of recommendation is different for each organization, and is strongly linked to its mission. 

\subsubsection{News}
Both the News organization and the Broadcaster highlight that they are ``for and from everyone". For the News organization, this is fairly straightforward: to \emph{``reach the widest possible audience, [and] enable people to be well-informed"} (\gerard). This means that the news should be accessible to anyone, regardless of skill level, and that journalistic quality takes precedence over personal interest (\tom,\gerard). Each article has a non-personalized `more like this' section which is automatically populated by a recommender system, but can be supplemented or turned off by the editorial team. Furthermore, there is a personalized recommender system under development that aims to connect readers to ``important news they have missed" (\tom). For this, the organization is experimenting with how behavioral data and editorial selections can complement each other, in order to compete against the commercial players (\tom). 

The News organization notes a very strong collaboration between the technical and editorial teams. When opening the app, the users will always land on the editorially curated front page listing the most important news of that moment. At any time, the editorial team has the power to turn off the recommender systems.
While diversity is an important concept in the news organization, it is not something that is currently explicitly built into the recommender system. Rather, it is seen as a procedural thing, to be considered at every step of content creation. This goes from choosing which stories and events to cover, to which people are doing the reporting, to writing about events in a neutral way covering multiple perspectives. 
This control results in a set of items to recommend from that \emph{``have to be told from a diverse standpoint in the first [place]"} (\tom). The organization sees the UX design of the recommender system, including where it is placed within the app and accompanied by which headers and explanations as more impactful (\tom). They do see potential in personalization of style, telling the news through the user's desired medium (text, video, audio) or in a language complexity level suitable to the reader (\gerard). 
\subsubsection{Broadcaster}
For the Broadcaster, the mission to be `from and for everyone' translates a little differently. The organization is effectively an umbrella for multiple smaller broadcasters, each representing a different section of Dutch society. They are the ones pitching ideas and creating the content, though the public broadcaster may request a certain type of program if they feel a particular perspective is missing (\jos). The broadcaster then tries to balance on the one hand helping their users recognize themselves in the content they have on offer, while also fostering understanding and knowledge of people, ideas and groups that are `different' (\susanne,\mezen,\bob,\tycho). For this, they see a clear purpose for personalized and diverse recommendation: \emph{``[I]f someone believes or thinks or feels in X, and they look at our platform, that person is also confronted with Y. And that Y is slightly different from what the person had in mind with X.''} (\mezen). Diversity plays a large role in achieving this, but the split in goals between recognition and broadening causes a great deal of conceptual unclarity. As one of the interviewees notes, \emph{``[i]f you are looking at personalization, then we don't want people to all get the same thing recommended. That's what I tend to see as diversity. [...] [P]luralism in recommendations would be [that] we'd like to look into [a] topic from different perspectives. And I think if you talk to different people within this organisation, those two things kind of mix.''} (\arno).  

The current platform is largely manually curated, with separate sections (displayed relatively low on the landing page) for algorithmic recommendations. These algorithmic recommendations balance personal relevance, based on a user's past viewing behavior, with a so-called `public value score'. This score is an aggregate, obtained through a daily survey sent to users, in which they are asked to score the programs they watched recently on things such as the presence of certain population groups or multiple perspectives. The broadcaster is working on the development of a new platform, which should have a much stronger algorithmic focus. They hold a uniformly strong position on user control: while they, as the broadcaster, should provide a diverse offering, the user is eventually always in charge of what they do and do not watch.

\subsubsection{Library}

The Library hosts a vast collection containing every book published in or about the Netherlands (\michel).
They hold a unique position among the cultural institutes, as they have the added role of coordinating 138 other public libraries.
There is high interest in digitization and innovation in general, which manifests in creation of proof-of-concept applications that the other libraries can choose to adopt or not (\mirjam).
Among other services and initiatives, the Library has the dual goal of having \textit{more people read} and \textit{people read more} (\lennie). 
The Library suspects that the decline of reading among its users is in part caused by a general lack of available time. 
Currently, there is a recommender system feature in the mobile application hosted by the Library, but its functionality is not satisfactory as \emph{``people cannot find something they might be interested in''} (\lennie), an issue partly caused by the organization of metadata.
Improved personalization can assist with the goal of attracting more patrons, as it increases accessibility to the collection for different types of people. 
At the same time, the Library is conducting research on the ethics around recommender systems, with topics like bias and privacy being central (\mirjam).
In this sense, personalization is not sufficient. 
There are active efforts to reduce bias that historically existed in the collections, in order to facilitate \emph{``every citizen to be able to participate in society'' and ``make society [as a whole] better, smarter and more creative''} (\michel).


Overall, the Library aims to deploy a well-functioning recommender system to satisfy their readers while remaining inclusive and enacting bias correction.
However, building such a system is not trivial, and there are many questions about how development should be approached.

\subsection{Aspects of diversity}
\label{sec:aspects}

\begin{figure*}[htp!]
\includegraphics[width=\textwidth]{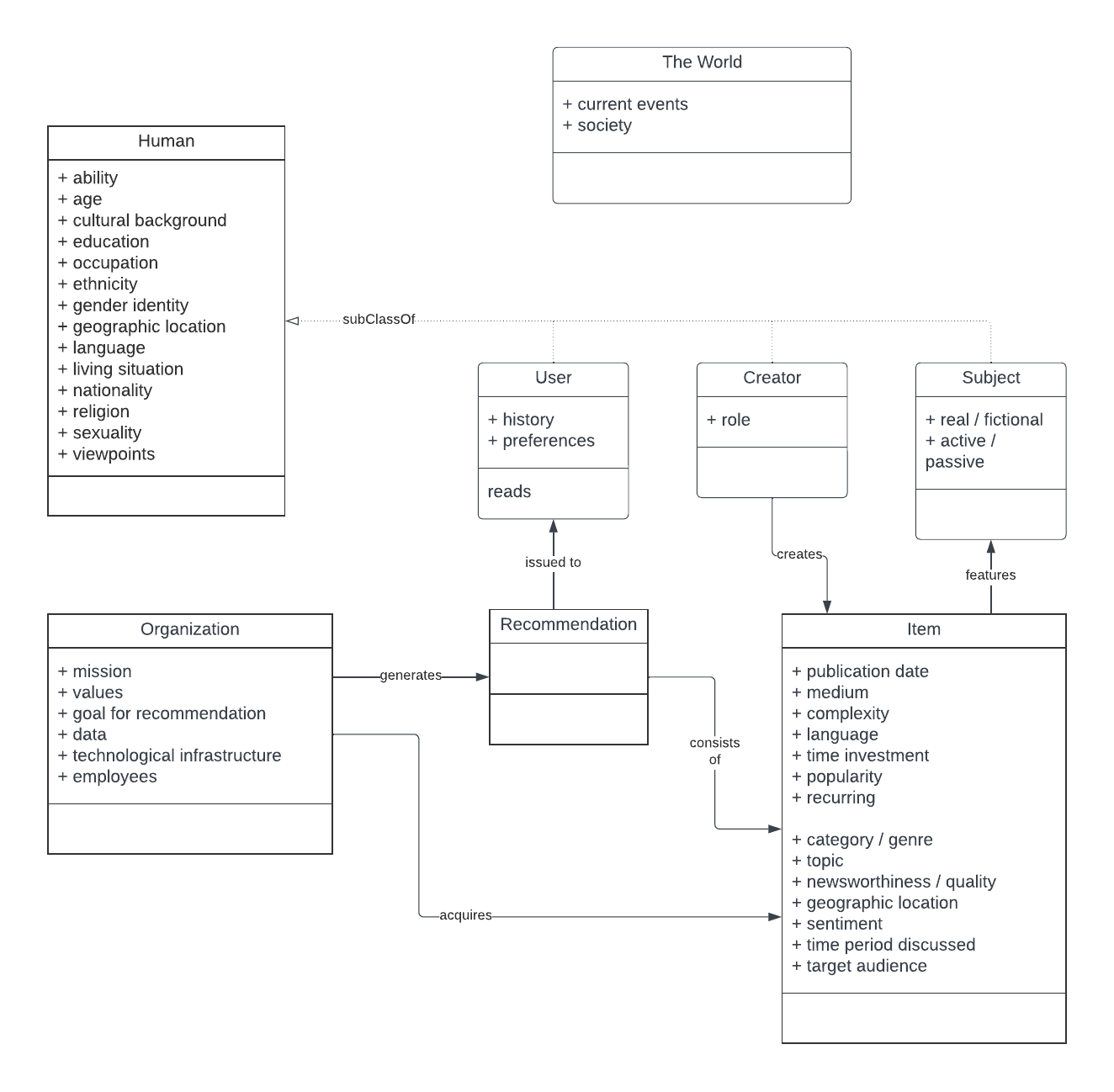}
\caption{Schematic overview of the identified aspects of diversity and how they interact with each other. The 'World' class is unconnected in the graph, but in truth encapsulates and informs everything: from the content that is being produced, to what user wants to read, to what constitutes a `viewpoint' or a `minority'.}
\label{fig:overview}
\Description{Schematic overview of the identified aspects of diversity and how they interact with each other. Aspects are divided in classes `Item', `Human' and `World'. Human has subclasses `User', `Creator' and `Subject'.}
\end{figure*}

During the recommendation exercise, and the questions leading up to it, participants would mention aspects of the content or the user beyond personal relevance that would lead them to include that item in the diverse recommendation or not. Figure \ref{fig:overview} provides a schematic overview of these aspects, which can be divided into Item-, Human- and World aspects. 

\subsubsection{Item aspects}
\label{sec:aspect_item}
The first dimension considers aspects of the items that are to be recommended. For the Library these would be books, news articles for the news agency, and videos for the Broadcaster. The first group of aspects revolves around \emph{topicality}, or what is actually discussed in the items. The 
dimensions mentioned by most interviewees are \textbf{category}, or sometimes genre, and \textbf{topic}. All interviewees but one mentioned balancing different categories and topics in the recommendation, to avoid saturation and to keep a user engaged. This is reflective of how diversity is currently most often conceptualized in technical recommender system literature; see \cite{castells2021novelty}. 

Related but still separate from the topic are the \textbf{geographic location} the item centers on, and the \textbf{time period} it discusses. The Library may want to recommend books that discuss a topic through the lens of different time periods (\janwillem), whereas the News organization aims to ensure that they not only cover news from densely populated urban areas, but also from rural areas (\gerard). Secondly, recommendations may vary on accounts of \emph{stylistic} properties. This relates to the way in which a message is communicated, and whether it is easily accessible for the user. Examples are the \textbf{complexity} of the content, the amount of \textbf{time investment} a user is required (and willing) to make, the \textbf{language} it is written in, the \textbf{target audience} of the item, and its \textbf{medium} (text, video, audio). 
Many of these properties are symmetric with characteristics of the user: a certain item complexity level or language requires a certain amount of skill from the user. 

Thirdly, there are a range of other item properties which may be prioritized. An important one is an item's \textbf{publication date}. While this is of primary importance to the news agency (one would not want to recommend a news item from two months ago), it is much less so for the broadcaster and the library: while there may be some preference for new publications, they also want to help their users discover less popular content. Another important dimension is the perceived \textbf{quality} of the item. While the news organization would explicitly optimize for (editorial) quality, opinions differ between participants from the library: while some would find it acceptable if readers only read comic books and manga, others wanted them to be pushed towards more high-brow literature (\lennie). More differences could be observed in terms of an item's \textbf{popularity}: some said that popular items are likely to be items that readers are looking for and should therefore be recommended (\lennie,\susanne), while others explicitly distanced themselves from it: \emph{``I do know that out of this selection, [sensationalist article] would have been clicked on more than [serious article]. [...] That's why what we're doing with those recommendations is designed the way we designed it, which is that the journalistic selection we make takes precedence."} (\gerard). 
Additionally, items that had a high production \textbf{cost} may be prioritized (\jos,\bob). 

Last, but definitely not least, interviewees referred to the people involved in the content. These can be split into the \textbf{Creators}, such as producers or authors, and the people that were \textbf{Subject} in the items, either through active participation or being discussed by others. Think of guests at a talkshow, politicians discussed in the news or a novel's protagonist. These are further discussed in Section \ref{sec:aspect_person}. 

\subsubsection{Human aspects}
\label{sec:aspect_person}
While item category is the 
aspect of diversity mentioned by most interviewees, it is closely followed by notions of the diversity in the context of \emph{people}. 
While there are a number of attributes that all humans share, there are three `subclasses' that can be derived with specific roles within the system. These are Users, Creators and Subjects. Users are the people that receive the recommendations, and have a set of unique properties that the other types do not have: their \textbf{history} and \textbf{preferences}. Creators are, as the name implies, in some way or another involved in the creation of the items; the authors of books, journalists or producers of shows. Lastly, there are the people that are the Subject of the item's content, such as protagonists of books or people appearing on talk shows. These subjects can be fictional or real, and be active or passive agents within the Item by either speaking on their own behalf or by being described or mentioned by others.    

The general Human aspects, such as \textbf{age}, \textbf{gender identity}, etc, are applicable to each type. A frequently mentioned aspect of diversity is relating to a person's background. \textbf{Culture}, \textbf{ethnicity}, \textbf{nationality} and sometimes \textbf{religion} (for example `Muslims') are often used interchangeably, and the distinction or relation between them is not always clear. For example, \tycho~notes: \emph{``I think about [...] representing different societal groups. So ethnic minorities. More black and white. Maybe a bit more foreign language [...]"}, while \michel~says: \emph{`` I don't know if this persona is from a specific country or has a specific background but [author] is an American author."} 
Identifying the presence of different ethnic and/or cultural groups is notoriously difficult, partly
because the data required to do so is often lacking. In each of the domains described above, textual data was the only type of data available, and often no additional information was present beside a person's name\footnote{Some approaches have attempted to predict a person's ethnicity based on their name, but this process is not generally accurate and can suffer from misrecognition bias \cite{lockhart2023name}}. In some cases, data enrichment might be possible (\jos). 
Yet, this requires building a dataset on attributes that are often considered sensitive, which gets even more problematic when considering the fact that, to `expand horizons' or `represent', a similar type of dataset would be necessary about the users, which would then lead to issues of privacy (\michel); see also \citet{papageorgiou2023necessity}. Similar safety concerns can be raised for aspects like \textbf{gender identity} and \textbf{sexuality}; see \citet{pinney2023much}.

\textbf{Culture}- and \textbf{viewpoint} diversity are sometimes mentioned in the same breath: \emph{``for me, diversity means having multiple colors, multiple opinions, multiple cultures, multiple points of view about something."} (\mezen). We make a distinction between the two, and denote viewpoint diversity as a person's perspective on events. Viewpoint diversity then often becomes linked to politics, or opinions about current events. A rich body of work exists around so-called viewpoint or stance detection, aiming to algorithmically extract these from text~\cite{draws2022comprehensive,aldayel2021stance,rieger2021item}. Even when successful, it is often unclear what a diversity of viewpoints would look effectively like, which is further discussed in Section \ref{sec:tactics}. 

\subsubsection{World aspects}
\label{sec:aspect_world}
Aspects related to the World are beyond the direct control of a recommender system or user. They influence which content is created, and as such the pool of items that a recommender system can make its selection from. They also interact with a user's preferences to determine what type of content a user is interested in, looking for, or should be reading. \textbf{Current events} determine what is newsworthy, and what news a user needs to consume to fulfill their information needs. Naturally this aspect is of high importance to the News organization (\tom,\gerard) and to some extent the Broadcaster (\jos, \tycho), but much less so to the Library. In contrast, \textbf{society} represents the world as it currently is, including its biases and existing power distributions. An organization may choose to either reflect or counter these existing structures (see Section \ref{sec:tactics_world}). For example, the Library acknowledges that much of their catalogue consists of white, American male authors, and wants to account for this in their recommendations (\janwillem).

\subsection{Tactics for diversification}
\label{sec:tactics}
During the experiment, the participants considered and deployed different tactics to produce a diverse recommendation.
Some participants articulated the tactic they deployed explicitly, while others did not clarify it in as much detail. 

\subsubsection{Diversity between items} 

Diversity between items is perhaps the most intuitive interpretation of diversity, and refers to ensuring that all items within a list of recommendations are sufficiently different between them.
Following this tactic requires choosing one or more appropriate axes to diversify over. 
In that sense, constructing a diverse list also entails exclusion, since some aspects have to be deemed less relevant.
This issue is particularly important for public organizations; selecting meaningful aspects for which diversity needs to be safeguarded is a crucial decision that must comply with the values and mission of the organization. Multiple interviewees constructed a recommended list by considering diversity between items in the process (\mezen,\jos,\bob,\tycho,\arno,\gerard,\janwillem,\michel,\mirjam,\lennie), especially to recommend items that are diverse between them in terms of category. 
In particular, the Broadcaster may aim for a \emph{``balance between [e]ntertainment and information''} (\mezen), given their wide range of offerings and mission \emph{``to inform, educate, and entertain''} (\mezen).
For the Library, item category translates to book genre, and is commonly taken into account when creating a diverse list (\michel,\mirjam,\lennie). 

Instead of generally diversifying between items, a somewhat adapted tactic is to recommend a set of items that are different between them in some perspective, but engage with the same topic or theme. 
For example, one interviewee opted for selecting \emph{``five books that give [...] an insight on [...] LGBTQ [...] [ ,h]ow that works or is around the world in different cultures and different times''} (\janwillem).
Furthermore, in case of a socially relevant emerging topic, the Broadcaster can create \emph{``a highlight lane and then offer a lot of different opinions that people can scroll through''} (\tycho).


Finally, interviewees noted as a result of the experiment that \emph{``to combine several aspects [is] really hard''} (\michel). 
It might be that a set of items is diverse over one important axis, but not over a different equally important one. 
In this context the act of creating a diverse recommendation can be seen as an optimization process that an algorithm can contribute to.
Regardless, ensuring diversity between items requires some sort of aspect prioritization, as well as an appropriate justification for it.







\subsubsection{Diversity as a within-item measure} 
A different tactic for ensuring diversity is recommending items that consist of diverse perspectives or types of people within the item itself. 
For example, according to a participant from the Broadcaster, a travel series that allows the viewer to \emph{``see multiple worlds [...] fits very well into diversity''} (\mezen).
This also pertains to episodes of political programs where \emph{``people from a lot of the major parties [appear], which captures a big part of the political spectrum''} (\bob).
The News organization notes that their inventory consists of \emph{``very good stories which have to be told from a diverse standpoint in the first [place]''} (\tom).
From this perspective, guaranteeing item diversity may be a part of the production/acquisition process or an explicit step of the recommendation.  
During the exercise, one interviewee (\bob) suggested combining within-item and between items, by composing a list of items on different topics where each item contains multiple perspectives on the respective topic. 
According to the interviewee, pursuing diversity in recommendation can also be seen as a process of achieving maximum diversity. 
This tactic requires that the media organization's catalogue consists of items that can facilitate the maximization process. 

\subsubsection{Diversity considering the user} 
\label{sec:tactics_user}

User-specific diversity is a personalized form of diversity, where the user's history and/or preferences are taken into account when composing a diverse list of items to recommend.
The first way to operationalize this tactic is to cater to a user's specific needs that potentially diverge from the norm.
This can manifest in assisting the user with finding uniquely niche content.
Based on this outlook, diversity is about \emph{``ensuring that [...] everyone feels that there is something for [them]''} (\susanne) in the catalogue.
The Library and News organization also mentioned literacy in this context (\lennie,\tom), as \emph{``how [to] help those with difficulty reading [...] [is] also a diversity topic in a way''} (\tom).
Additionally, for the Library diversity in accordance with the user can help them \emph{``recognize themselves in the author or in the main characters or the topics [such as] age and physical ability and sexual orientation''} (\michel).
The second way to apply user-specific diversity is to recommend to a user content that extends further from their current interests, and that they might not be aware of. 
In other words, an organization can also recommend items so as to \emph{``give people [...] a broader perspective of what is available''} (\mirjam) and to \emph{``broaden the user's horizons''} (\tycho).
This tactic can be seen as a way to ensure that users do not \emph{``stay in their own bubble''} (\michel) by \emph{``educating people [that] there's more than [their] bubble''} (\tycho).

\subsubsection{Diversity considering the world} 
\label{sec:tactics_world}

Diversity considering the world entails recommending items that in some way reflect or diverge from the norms that exist in the world, leading to `similar to world' and `different from world' diversification tactics. 
When reflecting the world, one could think of having a good spread of topics that are representative of the important news of that day (\gerard), or, by representing political parties in a way similar to their distribution in government (\bob). 
With a `different from world' diversification tactic, the aim is to counter existing power structures.
With this interpretation, an item can be considered diverse on its own merit.
This can be because the content represents a minority culture, an ``outsider [...] in a political sense" (\bob), a \emph{``very different part of the world that [the society] knows too little about''} (\mezen), or even a theme or topic that \emph{``you do not find that many [items] in the catalogue''} (\lennie).
In this case, very popular items may be deemed inherently not diverse, and might not need to receive further exposure: \emph{``[I]t's going to be on the website probably anyway''} (\arno).

What constitutes a minority was often implicitly assumed by the interviewees, potentially due to the social context that their organization operates in.
For example, multiple interviewees referred to LGBTQ-themed media as inherently diverse (\bob,\tycho,\arno,\michel), which also prompted their inclusion in a diverse recommendation as part of the experiment.
The same can be said for media that gives exposure to people with a minority cultural background or engages with topics not related to the Randstad, a dominant urban area in the Netherlands (\gerard).
From this perspective, diversity considering the world and user-based diversity (in a broadening-horizons way) can overlap when a user is assumed to be the typical or default representation of the majority culture in the given context.

\section{Discussion}
The wide range of interpretations highlighted in Section \ref{sec:results} show that, even within this limited domain, conceptualizations of diversity among participants vary greatly. 
It is therefore unlikely that a standardized definition, encompassing all potential goals, aspects and tactics, can be attained; however, that does not mean that it is impossible to implement a meaningful form of diversity into a recommender system. 
In this section, we outline the implications the findings of this study have for the implementation of diversity-aware recommender systems.

\subsection{Implementing diversity: a normative process}
Diversity has traditionally been seen as something we always want \emph{more} of: more viewpoints, more topics, more different nationalities. However, during the interview participants often mentioned instances in which they would actually want \emph{less}. An interviewee from the Library wanted to recommend items that fit the amount of time a user was willing to invest, while the News organization only wanted to recommend items of a certain quality. 
In these cases, meaningful diversity could be achieved by controlling for one aspect and diversifying over another; for example, when recommending different viewpoints for one controlled topic, or vice versa, by showing the opinions of one particular group over a wide range of topics. The same dynamic also occurs in the way interviewees spoke about diversification tactics. In some cases, a recommendation similar to a user's preferences or history would be desirable, in order for them to recognize themselves in the content on offer. In others, it should be different, so as to expose them to new perspectives. 
This shows that different applications may not only prioritize different aspects of diversity, they may also have different expectations on whether a recommendation should have high or low diversity.

Participants would often mention that recommendations should be `diverse enough.' This requires an underlying model that not only determines which items are similar and different, but also informs the system whether sufficient diversity has been attained. This is non-trivial, especially in a domain such as news recommendation, where contexts change rapidly. One could imagine using an external source as a reference point: for example, the presence and size of political parties in government, or a country's composition in terms of cultural groups as determined by a national statistics agency. However, this yet again relies on a conscious decision of what is the `right' model to use and reflect through the recommendations, and each choice will have pros and cons. 
There is therefore a normative choice to be made about the type of diversity that is desired, with different aspects to consider, different diversification tactics, and different levels of diversity. The wide variety in the answers given by participants is also an indication that this normative choice is not one that can be taken lightly, and requires a good deal of internal discussion and alignment with all the relevant stakeholders involved 
before it can be satisfactorily modeled and implemented.

\subsection{Generalizing to domains beyond public service media}
While the results are not directly generalizable to other domains, there are likely elements of the aspects and tactics identified here that would also be applicable. For example, while music recommendation might put less stock in the diversity of people discussed in an item's content, they would be interested in the diversity of Creators~\cite{ferraro2021break}; similarly, while in-item diversity would not be applicable, they could be interested in countering existing biases through different-from-world recommendation tactics~\cite{dinnissen2023amplifying}. 
We believe that the goals, aspects and tactics of this paper can still serve as a starting point for discussion in other domains, which may make it easier to identify gaps and unique challenges. 

\subsection{Exploring versus defining diversity}
Participants varied greatly in which diversity aspects and diversification tactics they mentioned. Some of these differences can be traced to the different ways participants speak: some people are more verbose, more inclined to stick to a specific set of examples, or already have a more developed concept in mind than others. 
Furthermore, the three organizations were at different stages of developing strategies towards diversity, which may have an impact on said differences between participants. However, they are all working on or considering the implementation of a recommender system. Our participants are thus also actively making decisions about how diversity would be conceptualized and implemented. They are as such representative of the `real' world, rather than the `ideal' world, and therefore suitable candidates for an exploration of the dimensions along which diversity is currently conceptualized. For these reasons, we refrain from making claims about whether certain aspects or tactics are `more important' than others, nor do we make claims about the `correct' definition of diversity. 
Our hope is that the overviews of goals, aspects and tactics presented in this study will facilitate building a common understanding and vocabulary between stakeholders with a different background, making it easier to find common ground and establish priorities that reflect the requirements of that particular implementation.

\section{Conclusion}
Through interviews with participants from public service media organizations, we identified a range of different interpretations of diversity within a recommender system. We grouped these into different goals (e.g. broadly informing the public or allowing a user to recognize themselves), aspects (e.g. the topic of the content, or the cultural background of an author), and diversification tactics (e.g. ensuring  diversity within a single item or countering biases in society). 
Given the great variety found in the conceptualization of diversity in recommender systems, even within a limited domain, we find that it is unlikely that a standardized definition can be attained. 
Instead, rather than striving for this standardization, we argue that it should be conceptualized on a case-by-case basis.
We hope that our mapping of goals, aspects, and tactics can contribute in effectively communicating what diversity entails within a specific application.  

\begin{acks}
This work was supported with seed funding from the RPA Human(e) AI, round 2022/2023, and the AI, Media and Democracy Lab, NWA.1332.20.009. The authors would like to express their gratitude to everyone that provided input and feedback on the project; Sophie Morosoli and Hannes Cools for their help with the methodology; Naomi Appelman, Kimon Kieslich, Midas Nouwens and Marijn Sax,  for providing input that helped shape the direction of the paper; the three anonymous FAccT reviewers for their helpful feedback; and lastly to Natali Helberger and Claes de Vreese for their support and reviews.
\end{acks}


\section*{Statements}
\subsection*{Ethical considerations statement}
A main ethical concern in interview-based research is breaching the anonymity of the interviewees. To support anonymity, we refer to the organizations by their general functionality instead of naming them. To the interviewees themselves we refer by their organization's functionality as well as a random identifier. Even though we did consider their position in the organization when selecting them, we do not include it when quoting them, as an extra measure for anonymity. Additionally, we asked the participants to sign an informed consent document. We explicitly asked for permission to record the interviews. Finally, we forwarded to them the conclusions of our research and quotes attributed to them to ensure we correctly interpreted and contextualized their words.
\subsection*{Researcher positionality statement}
The authors of this work are all European citizens who operate in the academic circles of the Netherlands. This shaped the work as we interviewed practitioners from the Netherlands who all work for organizations that we are familiar with in a professional and personal capacity, and some are in turn partly familiar with our work, which rendered communication with them easier. Three out of the four authors of this paper work in the computer science field, and one has joint expertise in computer science and communication science. For this reason we also received help (see Acknowledgments) from communication scientists, in particular when it comes to structuring the interview and devising a coding scheme. One author works closely with cultural institutes, one with media institutes, and one with both, which helps us contextualize the statements of the interviewees. 
\subsection*{Adverse impact statement}
Despite our efforts to ensure anonymity of the individual interviewees, it might be that readers familiar with the media landscape of the Netherlands can deduce which organizations we refer to in this paper. 
Additionally, readers of this paper might generalize the personal statements of the interviewees as completely representative of the entire organization that they work in.
Finally, we do not intend our mapping to serve as a final and conclusive categorization of media diversity, but it might be interpreted as such by readers.

\bibliographystyle{ACM-Reference-Format}
\bibliography{references}

\newpage
\onecolumn
\title{Appendix}
\begin{appendices}

\section{Coding scheme and mentions}
\label{img:mentions}
\begin{table*}[htp]
\caption{Final coding scheme, containing both aspects and tactics. Aspects are divided into Item, Person and World aspects. }
\label{tab:mentions}
\resizebox{0.95\textwidth}{!}{%
\begin{tabular}{clllllllllllll}
\rowcolor[HTML]{656565} 
\multicolumn{1}{l}{\cellcolor[HTML]{656565}{\color[HTML]{FFFFFF} \textbf{}}} &
  {\color[HTML]{FFFFFF} \textbf{}} &
  {\color[HTML]{FFFFFF} \textbf{B1}} &
  {\color[HTML]{FFFFFF} \textbf{B2}} &
  {\color[HTML]{FFFFFF} \textbf{B3}} &
  {\color[HTML]{FFFFFF} \textbf{B4}} &
  {\color[HTML]{FFFFFF} \textbf{B5}} &
  {\color[HTML]{FFFFFF} \textbf{B6}} &
  {\color[HTML]{FFFFFF} \textbf{L1}} &
  {\color[HTML]{FFFFFF} \textbf{L2}} &
  {\color[HTML]{FFFFFF} \textbf{L3}} &
  {\color[HTML]{FFFFFF} \textbf{L4}} &
  {\color[HTML]{FFFFFF} \textbf{N1}} &
  {\color[HTML]{FFFFFF} \textbf{N2}} \\ \hline
\multicolumn{1}{c|}{} &
  \multicolumn{1}{l|}{category / genre} &
  \cellcolor[HTML]{F2F2F2}x &
  \cellcolor[HTML]{F2F2F2}x &
  \cellcolor[HTML]{F2F2F2}x &
  \cellcolor[HTML]{F2F2F2}x &
  \cellcolor[HTML]{F2F2F2}x &
  \cellcolor[HTML]{F2F2F2}x &
  \cellcolor[HTML]{F2F2F2}x &
  \cellcolor[HTML]{F2F2F2}x &
  \cellcolor[HTML]{F2F2F2}x &
  \cellcolor[HTML]{F2F2F2}x &
  \cellcolor[HTML]{F2F2F2}x &
   \\
\multicolumn{1}{c|}{} &
  \multicolumn{1}{l|}{complexity} &
   &
  \cellcolor[HTML]{F2F2F2}x &
  \cellcolor[HTML]{F2F2F2}x &
  \cellcolor[HTML]{F2F2F2}x &
   &
   &
   &
   &
   &
  \cellcolor[HTML]{F2F2F2}x &
  \cellcolor[HTML]{F2F2F2}x &
   \\
\multicolumn{1}{c|}{} &
  \multicolumn{1}{l|}{cost} &
   &
   &
   &
  \cellcolor[HTML]{F2F2F2}x &
   &
   &
   &
   &
   &
   &
   &
   \\
\multicolumn{1}{c|}{} &
  \multicolumn{1}{l|}{creator (person)} &
  \cellcolor[HTML]{F2F2F2}x &
   &
  \cellcolor[HTML]{F2F2F2}x &
  \cellcolor[HTML]{F2F2F2}x &
  \cellcolor[HTML]{F2F2F2}x &
   &
  \cellcolor[HTML]{F2F2F2}x &
  \cellcolor[HTML]{F2F2F2}x &
  \cellcolor[HTML]{F2F2F2}x &
  \cellcolor[HTML]{F2F2F2}x &
   &
   \\
\multicolumn{1}{c|}{} &
  \multicolumn{1}{l|}{geographic location} &
  \cellcolor[HTML]{F2F2F2}x &
  \cellcolor[HTML]{F2F2F2}x &
  \cellcolor[HTML]{F2F2F2}x &
  \cellcolor[HTML]{F2F2F2}x &
  \cellcolor[HTML]{F2F2F2}x &
  \cellcolor[HTML]{F2F2F2}x &
  \cellcolor[HTML]{F2F2F2}x &
   &
  \cellcolor[HTML]{F2F2F2}x &
   &
  \cellcolor[HTML]{F2F2F2}x &
  \cellcolor[HTML]{F2F2F2}x \\
\multicolumn{1}{c|}{} &
  \multicolumn{1}{l|}{language} &
   &
  \cellcolor[HTML]{F2F2F2}x &
  \cellcolor[HTML]{F2F2F2}x &
   &
   &
   &
  \cellcolor[HTML]{F2F2F2}x &
   &
   &
   &
   &
   \\
\multicolumn{1}{c|}{} &
  \multicolumn{1}{l|}{newsworthiness / quality} &
  \cellcolor[HTML]{F2F2F2}x &
  \cellcolor[HTML]{F2F2F2}x &
  \cellcolor[HTML]{F2F2F2}x &
  \cellcolor[HTML]{F2F2F2}x &
  \cellcolor[HTML]{F2F2F2}x &
   &
  \cellcolor[HTML]{F2F2F2}x &
   &
   &
  \cellcolor[HTML]{F2F2F2}x &
  \cellcolor[HTML]{F2F2F2}x &
  \cellcolor[HTML]{F2F2F2}x \\
\multicolumn{1}{c|}{} &
  \multicolumn{1}{l|}{medium} &
   &
   &
   &
   &
   &
   &
   &
   &
   &
   &
   &
  \cellcolor[HTML]{F2F2F2}x \\
\multicolumn{1}{c|}{} &
  \multicolumn{1}{l|}{subject (person)} &
  \cellcolor[HTML]{F2F2F2}x &
  \cellcolor[HTML]{F2F2F2}x &
  \cellcolor[HTML]{F2F2F2}x &
  \cellcolor[HTML]{F2F2F2}x &
  \cellcolor[HTML]{F2F2F2}x &
  \cellcolor[HTML]{F2F2F2}x &
   &
  \cellcolor[HTML]{F2F2F2}x &
   &
  \cellcolor[HTML]{F2F2F2}x &
   &
  \cellcolor[HTML]{F2F2F2}x \\
\multicolumn{1}{c|}{} &
  \multicolumn{1}{l|}{popularity} &
  \cellcolor[HTML]{F2F2F2}x &
  \cellcolor[HTML]{F2F2F2}x &
  \cellcolor[HTML]{F2F2F2}x &
  \cellcolor[HTML]{F2F2F2}x &
  \cellcolor[HTML]{F2F2F2}x &
  \cellcolor[HTML]{F2F2F2}x &
  \cellcolor[HTML]{F2F2F2}x &
   &
  \cellcolor[HTML]{F2F2F2}x &
   &
  \cellcolor[HTML]{F2F2F2}x &
  \cellcolor[HTML]{F2F2F2}x \\
\multicolumn{1}{c|}{} &
  \multicolumn{1}{l|}{publication date} &
  \cellcolor[HTML]{F2F2F2}x &
   &
  \cellcolor[HTML]{F2F2F2}x &
  \cellcolor[HTML]{F2F2F2}x &
   &
   &
  \cellcolor[HTML]{F2F2F2}x &
   &
   &
   &
  \cellcolor[HTML]{F2F2F2}x &
  \cellcolor[HTML]{F2F2F2}x \\
\multicolumn{1}{c|}{} &
  \multicolumn{1}{l|}{recurring} &
   &
   &
  \cellcolor[HTML]{F2F2F2}x &
   &
   &
   &
   &
   &
   &
   &
  \cellcolor[HTML]{F2F2F2}x &
   \\
\multicolumn{1}{c|}{} &
  \multicolumn{1}{l|}{relevance} &
  \cellcolor[HTML]{F2F2F2}x &
  \cellcolor[HTML]{F2F2F2}x &
  \cellcolor[HTML]{F2F2F2}x &
  \cellcolor[HTML]{F2F2F2}x &
  \cellcolor[HTML]{F2F2F2}x &
   &
  \cellcolor[HTML]{F2F2F2}x &
   &
  \cellcolor[HTML]{F2F2F2}x &
   &
  \cellcolor[HTML]{F2F2F2}x &
  \cellcolor[HTML]{F2F2F2}x \\
\multicolumn{1}{c|}{} &
  \multicolumn{1}{l|}{sentiment} &
   &
   &
   &
   &
  \cellcolor[HTML]{F2F2F2}x &
   &
   &
   &
   &
   &
   &
   \\
\multicolumn{1}{c|}{} &
  \multicolumn{1}{l|}{target audience} &
   &
  \cellcolor[HTML]{F2F2F2}x &
  \cellcolor[HTML]{F2F2F2}x &
   &
   &
   &
   &
   &
   &
   &
   &
  \cellcolor[HTML]{F2F2F2}x \\
\multicolumn{1}{c|}{} &
  \multicolumn{1}{l|}{time investment} &
   &
   &
   &
   &
   &
   &
   &
   &
   &
  \cellcolor[HTML]{F2F2F2}x &
  \cellcolor[HTML]{F2F2F2}x &
   \\
\multicolumn{1}{c|}{} &
  \multicolumn{1}{l|}{time period discussed} &
   &
   &
   &
  \cellcolor[HTML]{F2F2F2}x &
  \cellcolor[HTML]{F2F2F2}x &
   &
  \cellcolor[HTML]{F2F2F2}x &
   &
   &
   &
   &
   \\
\multicolumn{1}{c|}{\multirow{-18}{*}{Item}} &
  \multicolumn{1}{l|}{topic} &
  \cellcolor[HTML]{F2F2F2}x &
  \cellcolor[HTML]{F2F2F2}x &
  \cellcolor[HTML]{F2F2F2}x &
  \cellcolor[HTML]{F2F2F2}x &
  \cellcolor[HTML]{F2F2F2}x &
  \cellcolor[HTML]{F2F2F2}x &
  \cellcolor[HTML]{F2F2F2}x &
  \cellcolor[HTML]{F2F2F2}x &
  \cellcolor[HTML]{F2F2F2}x &
  \cellcolor[HTML]{F2F2F2}x &
  \cellcolor[HTML]{F2F2F2}x &
  \cellcolor[HTML]{F2F2F2}x \\ \hline
\multicolumn{1}{c|}{} &
  \multicolumn{1}{l|}{ability} &
   &
   &
   &
   &
   &
   &
   &
  \cellcolor[HTML]{F2F2F2}x &
  \cellcolor[HTML]{F2F2F2}x &
   &
   &
   \\
\multicolumn{1}{c|}{} &
  \multicolumn{1}{l|}{age} &
  \cellcolor[HTML]{F2F2F2}x &
  \cellcolor[HTML]{F2F2F2}x &
  \cellcolor[HTML]{F2F2F2}x &
  \cellcolor[HTML]{F2F2F2}x &
   &
   &
  \cellcolor[HTML]{F2F2F2}x &
  \cellcolor[HTML]{F2F2F2}x &
  \cellcolor[HTML]{F2F2F2}x &
   &
   &
   \\
\multicolumn{1}{c|}{} &
  \multicolumn{1}{l|}{cultural background} &
   &
  \cellcolor[HTML]{F2F2F2}x &
   &
  \cellcolor[HTML]{F2F2F2}x &
  \cellcolor[HTML]{F2F2F2}x &
  \cellcolor[HTML]{F2F2F2}x &
  \cellcolor[HTML]{F2F2F2}x &
  \cellcolor[HTML]{F2F2F2}x &
   &
   &
   &
   \\
\multicolumn{1}{c|}{} &
  \multicolumn{1}{l|}{education} &
   &
   &
   &
  \cellcolor[HTML]{F2F2F2}x &
   &
   &
   &
   &
   &
   &
  \cellcolor[HTML]{F2F2F2}x &
  \cellcolor[HTML]{F2F2F2}x \\
\multicolumn{1}{c|}{} &
  \multicolumn{1}{l|}{ethnicity} &
   &
  \cellcolor[HTML]{F2F2F2}x &
  \cellcolor[HTML]{F2F2F2}x &
  \cellcolor[HTML]{F2F2F2}x &
   &
   &
   &
  \cellcolor[HTML]{F2F2F2}x &
   &
   &
   &
   \\
\multicolumn{1}{c|}{} &
  \multicolumn{1}{l|}{gender identity} &
   &
  \cellcolor[HTML]{F2F2F2}x &
  \cellcolor[HTML]{F2F2F2}x &
  \cellcolor[HTML]{F2F2F2}x &
  \cellcolor[HTML]{F2F2F2}x &
   &
  \cellcolor[HTML]{F2F2F2}x &
  \cellcolor[HTML]{F2F2F2}x &
  \cellcolor[HTML]{F2F2F2}x &
   &
   &
   \\
\multicolumn{1}{c|}{} &
  \multicolumn{1}{l|}{geographic location} &
  \cellcolor[HTML]{F2F2F2}x &
  \cellcolor[HTML]{F2F2F2}x &
  \cellcolor[HTML]{F2F2F2}x &
  \cellcolor[HTML]{F2F2F2}x &
   &
  \cellcolor[HTML]{F2F2F2}x &
  \cellcolor[HTML]{F2F2F2}x &
   &
  \cellcolor[HTML]{F2F2F2}x &
   &
  \cellcolor[HTML]{F2F2F2}x &
   \\
\multicolumn{1}{c|}{} &
  \multicolumn{1}{l|}{living situation} &
   &
   &
  \cellcolor[HTML]{F2F2F2}x &
  \cellcolor[HTML]{F2F2F2}x &
  \cellcolor[HTML]{F2F2F2}x &
  \cellcolor[HTML]{F2F2F2}x &
   &
  \cellcolor[HTML]{F2F2F2}x &
   &
   &
  \cellcolor[HTML]{F2F2F2}x &
   \\
\multicolumn{1}{c|}{} &
  \multicolumn{1}{l|}{nationality} &
   &
  \cellcolor[HTML]{F2F2F2}x &
  \cellcolor[HTML]{F2F2F2}x &
  \cellcolor[HTML]{F2F2F2}x &
  \cellcolor[HTML]{F2F2F2}x &
   &
   &
  \cellcolor[HTML]{F2F2F2}x &
  \cellcolor[HTML]{F2F2F2}x &
  \cellcolor[HTML]{F2F2F2}x &
  \cellcolor[HTML]{F2F2F2}x &
   \\
\multicolumn{1}{c|}{} &
  \multicolumn{1}{l|}{religion} &
  \cellcolor[HTML]{F2F2F2}x &
  \cellcolor[HTML]{F2F2F2}x &
  \cellcolor[HTML]{F2F2F2}x &
  \cellcolor[HTML]{F2F2F2}x &
  \cellcolor[HTML]{F2F2F2}x &
  \cellcolor[HTML]{F2F2F2}x &
   &
   &
  \cellcolor[HTML]{F2F2F2}x &
   &
   &
   \\
\multicolumn{1}{c|}{} &
  \multicolumn{1}{l|}{sexuality} &
  \cellcolor[HTML]{F2F2F2}x &
   &
   &
  \cellcolor[HTML]{F2F2F2}x &
  \cellcolor[HTML]{F2F2F2}x &
  \cellcolor[HTML]{F2F2F2}x &
  \cellcolor[HTML]{F2F2F2}x &
  \cellcolor[HTML]{F2F2F2}x &
  \cellcolor[HTML]{F2F2F2}x &
  \cellcolor[HTML]{F2F2F2}x &
   &
   \\
\multicolumn{1}{c|}{} &
  \multicolumn{1}{l|}{viewpoint} &
  \cellcolor[HTML]{F2F2F2}x &
  \cellcolor[HTML]{F2F2F2}x &
  \cellcolor[HTML]{F2F2F2}x &
  \cellcolor[HTML]{F2F2F2}x &
  \cellcolor[HTML]{F2F2F2}x &
  \cellcolor[HTML]{F2F2F2}x &
  \cellcolor[HTML]{F2F2F2}x &
  \cellcolor[HTML]{F2F2F2}x &
  \cellcolor[HTML]{F2F2F2}x &
   &
  \cellcolor[HTML]{F2F2F2}x &
   \\
\multicolumn{1}{c|}{} &
  \multicolumn{1}{l|}{user; history} &
  \cellcolor[HTML]{F2F2F2}x &
  \cellcolor[HTML]{F2F2F2}x &
  \cellcolor[HTML]{F2F2F2}x &
  \cellcolor[HTML]{F2F2F2}x &
   &
   &
   &
  \cellcolor[HTML]{F2F2F2}x &
  \cellcolor[HTML]{F2F2F2}x &
   &
  \cellcolor[HTML]{F2F2F2}x &
   \\
\multicolumn{1}{c|}{\multirow{-14}{*}{Person}} &
  \multicolumn{1}{l|}{user; preferences} &
  \cellcolor[HTML]{F2F2F2}x &
   &
  \cellcolor[HTML]{F2F2F2}x &
  \cellcolor[HTML]{F2F2F2}x &
  \cellcolor[HTML]{F2F2F2}x &
   &
   &
  \cellcolor[HTML]{F2F2F2}x &
  \cellcolor[HTML]{F2F2F2}x &
   &
  \cellcolor[HTML]{F2F2F2}x &
   \\ \hline
\multicolumn{1}{c|}{} &
  \multicolumn{1}{l|}{current events} &
   &
   &
  \cellcolor[HTML]{F2F2F2}x &
   &
  \cellcolor[HTML]{F2F2F2}x &
   &
   &
   &
   &
   &
  \cellcolor[HTML]{F2F2F2}x &
  \cellcolor[HTML]{F2F2F2}x \\
\multicolumn{1}{c|}{\multirow{-2}{*}{World}} &
  \multicolumn{1}{l|}{society} &
   &
   &
   &
  \cellcolor[HTML]{F2F2F2}x &
  \cellcolor[HTML]{F2F2F2}x &
   &
   &
   &
  \cellcolor[HTML]{F2F2F2}x &
   &
  \cellcolor[HTML]{F2F2F2}x &
   \\
\rowcolor[HTML]{656565} 
\multicolumn{14}{l}{\cellcolor[HTML]{656565}{\color[HTML]{FFFFFF} \textbf{Tactics}}} \\
\multicolumn{1}{c|}{} &
  \multicolumn{1}{l|}{different in item} &
   &
  \cellcolor[HTML]{F2F2F2}x &
   &
  \cellcolor[HTML]{F2F2F2}x &
  \cellcolor[HTML]{F2F2F2}x &
  \cellcolor[HTML]{F2F2F2}x &
   &
   &
   &
   &
  \cellcolor[HTML]{F2F2F2}x &
   \\
\multicolumn{1}{c|}{} &
  \multicolumn{1}{l|}{different in list} &
  \cellcolor[HTML]{F2F2F2}x &
  \cellcolor[HTML]{F2F2F2}x &
  \cellcolor[HTML]{F2F2F2}x &
  \cellcolor[HTML]{F2F2F2}x &
  \cellcolor[HTML]{F2F2F2}x &
  \cellcolor[HTML]{F2F2F2}x &
  \cellcolor[HTML]{F2F2F2}x &
  \cellcolor[HTML]{F2F2F2}x &
  \cellcolor[HTML]{F2F2F2}x &
  \cellcolor[HTML]{F2F2F2}x &
   &
  \cellcolor[HTML]{F2F2F2}x \\
\multicolumn{1}{c|}{} &
  \multicolumn{1}{l|}{similar to user} &
  \cellcolor[HTML]{F2F2F2}x &
   &
  \cellcolor[HTML]{F2F2F2}x &
   &
  \cellcolor[HTML]{F2F2F2}x &
  \cellcolor[HTML]{F2F2F2}x &
  \cellcolor[HTML]{F2F2F2}x &
  \cellcolor[HTML]{F2F2F2}x &
  \cellcolor[HTML]{F2F2F2}x &
  \cellcolor[HTML]{F2F2F2}x &
  \cellcolor[HTML]{F2F2F2}x &
  \cellcolor[HTML]{F2F2F2}x \\
\multicolumn{1}{c|}{} &
  \multicolumn{1}{l|}{different from user} &
  \cellcolor[HTML]{F2F2F2}x &
  \cellcolor[HTML]{F2F2F2}x &
   &
   &
  \cellcolor[HTML]{F2F2F2}x &
  \cellcolor[HTML]{F2F2F2}x &
   &
  \cellcolor[HTML]{F2F2F2}x &
  \cellcolor[HTML]{F2F2F2}x &
   &
   &
   \\
\multicolumn{1}{c|}{} &
  \multicolumn{1}{l|}{different/similar to world} &
   &
  \cellcolor[HTML]{F2F2F2}x &
   &
  \cellcolor[HTML]{F2F2F2}x &
  \cellcolor[HTML]{F2F2F2}x &
  \cellcolor[HTML]{F2F2F2}x &
  \cellcolor[HTML]{F2F2F2}x &
   &
   &
  \cellcolor[HTML]{F2F2F2}x &
   &
  
\end{tabular}%
}
\end{table*}


\end{appendices}

\end{document}